\documentstyle[]{mn}				
\input{epsf}					
\def \arcmin      { \prime }
\def \mpc       {{\rm\ Mpc}}

\def \yr        {{\rm\ yr}}

\def \eg        {\hbox{\it e.g.}}

\def \etal      {et al.\ }

\def \Ho        {{\rm\ H_{o}}}
\def \kmsmpc    {{\rm\ km\ s^{-1}\ Mpc^{-1}}}
\def \kev       {{\rm\ keV}}
\def \msol      {{\rm\ M}_\odot}

\def \hfifty    {\hbox{$\, h_{50}$} }
\def \hfiftyinv {\hbox{$\, h_{50}^{-1}$} }

\def \rms       {{\it rms~} }
\def \cgsflux   {{\rm\ erg\ s^{-1}\ cm^{-2}}}
\def \arcmin    {^\prime}

\def \se        {\!=\!}
\def \sims      {\sim \!}
\def \ssim      {\! \sim \!}
\def \ssimeq    {\! \simeq \!}
\def \sequiv    {\! \equiv \!}
\def \spropto   {\! \propto \!}
\def\\{\hfil\break}
\def\spose#1{\hbox to 0pt{#1\hss}}
\def\lta{\mathrel{\spose{\lower 3pt\hbox{$\mathchar"218$}}
     \raise 2.0pt\hbox{$\mathchar"13C$}}}
\def\gta{\mathrel{\spose{\lower 3pt\hbox{$\mathchar"218$}}
     \raise 2.0pt\hbox{$\mathchar"13E$}}}

\def\lesssim{\mathrel{\hbox{\rlap{\hbox{\lower4pt\hbox{$\sim$}}}\hbox{$<$}}}}
\def\gtrsim{\mathrel{\hbox{\rlap{\hbox{\lower4pt\hbox{$\sim$}}}\hbox{$>$}}}}

\newcount\itemno
\def \ino         { \the\itemno\global\advance\itemno by 1 }
\def\apj{ApJ}
\def\aj{AJ}

\def\aa{A\&A}
\def\mnras{MNRAS}
\def\araa{ARA\&A}
\def\nature{Nature}
\def\pasj{PASJ}

\def \xray {\hbox{X--ray} }
\def \rfiveh {\hbox{$r_{500}$}}

\def \deltac {\hbox{$\delta_c$}}
\def \rdeltac {\hbox{$r_{\delta_c}$}}
\def \Mdeltac {\hbox{$M_{\delta_c}$}}

\def \rhocrit {\hbox{$\rho_c$}}
\def \fbargas {\hbox{${\bar f}_{gas}$}}
\def \fgas {\hbox{$f_{gas}$}}

\def \LxT {\hbox{$L_X$--$T$} }
\def \betamodel {\hbox{$\beta$--model} }
\def \mdot {\hbox{$\dot{M}$} }

\def \logTd6 {\hbox{log$( T/6 \kev)$} }
\def \logLx {\hbox{log$(L_X)$} }
\def \fgas {\hbox{$f_{gas}$}}

\title[ Intracluster Gas Fractions of X--ray Clusters ]
{ The $L_X-T$ Relation and Intracluster Gas Fractions
of X--ray Clusters }
\author[M. Arnaud and A.E. Evrard]
{Monique Arnaud$^1$ and August E. Evrard$^{2,3}$ \\
   $^1$ C.E.A., DSM, DAPNIA, Service d'Astrophysique, C.E. Saclay,
   F-91191, Gif-Sur-Yvette Cedex, France \\
   $^2$ Institut d'Astrophysique, 98bis Blvd Arago, 75014 Paris, France \\
   $^3$ Physics Department, University of Michigan, Ann Arbor, MI
48109--1120 USA \\
arnaud@sapvxg.saclay.cea.fr \\
evrard@umich.edu }

\date{Accepted 1999 January 12}

\begin{document}

\maketitle

\begin{abstract}

We re-examine the \xray luminosity--temperature relation using a
nearly homogeneous data set of 24 clusters selected for statistically
accurate temperature measurements and absence of strong cooling flows.
The data exhibit a remarkably tight power--law relation between
bolometric luminosity and temperature with a slope $2.88 \pm
0.15$.  With reasonable assumptions regarding cluster structure, we
infer an upper limit on fractional variations in the intracluster gas
fraction $\langle (\delta \fgas / \fgas)^2 \rangle^{1/2} \le 15\%$.  A
strictly homogeneous GINGA subset of 18 clusters places a more
stringent limit of $9\%$.

Imaging data from the literature are employed to determine absolute
values of $\fgas$ within spheres encompassing density contrasts
$\delta_c \se 500$ and 200 with respect to the critical density.
Comparing binding mass estimates based on the virial theorem (VT) and
the hydrostatic, \betamodel (BM), we find a temperature--dependent
discrepancy in $\fgas$ between the two methods caused by systematic
variation of the outer slope parameter $\beta$ with temperature.  Mean
values (for $\Ho \se 50 \kmsmpc$) range from $\fbargas \se 0.10$ for
cool ($T < 4 \kev$) clusters using the VT at $\deltac \se 500$ to
$0.22$ for hot ($T > 4 \kev$) clusters using the BM at  $\deltac \se
200$.  There is evidence that cool clusters have a lower mean gas
fraction than hot clusters, but it is not possible to assess the
statistical significance of this effect in the present dataset.  The $T$
dependence of the ICM density structure, coupled with the increase of
the gas fraction with $T$ in the VT approach, explains the steepening
of the $\LxT$ relation.

The small variation about the mean gas fraction within this majority
sub--population of clusters presents an important constraint for
theories of galaxy formation and supports arguments against an
Einstein--deSitter universe based on the population mean gas fraction
and conventional, primordial nucleosynthesis.  The apparent trend of
lower gas fractions and more extended atmospheres in low
temperature systems are consistent with expectations of models
incorporating the effects of galactic winds on the intracluster
medium.

\end{abstract}


\section{Introduction}

The relation between \xray temperature $T$ and luminosity $L_X$ is a
sensitive diagnostic of structural regularity in clusters of galaxies.
Because the luminosity is governed by the mass of gas in the
intracluster medium (ICM) while the temperature is determined by the
total, gravitating cluster mass, the $\LxT$ relation can be used as a
tool to probe variations in gas fraction --- the ratio of ICM gas to
total mass --- in these systems.

Evidence for a correlation between these basic observational
quantities has existed since the early days of \xray astronomy
(Mitchell \etal 1977, 1979; Mushotzky \etal 1978; Edge \& Stewart
1991, David \etal 1993).  Despite its relative maturity, a key
characteristic of this relation was only recently established.  Fabian
\etal (1994) demonstrate that the departure of a cluster from the mean
$L_X-T$ relation is correlated with the strength of the emission
associated with its cooling flow core.  The picture resulting from
this work is of a mixed population which is likely to be time
variable.  Fabian \etal speculate that cooling flows are a recurrent
phenomenon associated with a secular instability in the ICM plasma
which periodically is interrupted and reset by strong merging
encounters.  Periodic mergers of varying strength are a natural
feature of hierarchical clustering models of structure formation, and
numerical simulations suggest that cooling flow features can be erased
by mergers (Garasi, Burns \& Loken 1997).
However, the theoretical picture remains incomplete because of the
complexities of the physics operating in the cluster core.  Regardless
of the exact physical mechanisms responsible, the empirical fact is
that the cluster population can be broadly classed into two categories
by making a cut in cooling flow strength $\mdot$.

In this paper, we present and analyze the luminosity--temperature
relation from a nearly homogeneous data set of weak cooling flow
clusters, where $\mdot \le 100 \msol \yr^{-1}$ defines weak.  The data
set is described in \S2, with particular attention paid to systematic
errors in determination of both $L_X$ and $T$.  In \S3, we present the
$\LxT$ relation and show that these data place stringent limits on gas
fraction variations within this sub--population.  Imaging data taken
from the literature allow us to estimate absolute values of the gas
fraction at fixed density contrast in \S4.  We compare results for two
different binding mass estimation methods, and emphasize the
systematic uncertainty introduced by this choice.

Hubble constant dependencies are displayed in the paper via $\hfifty
\equiv \Ho/50 \kmsmpc$ and we assume $\hfifty \se 1$ when quoting
numerical values throughout the paper.

\section{The Sample}

We analyse an archival data set, part of which was assembled and
discussed by Arnaud (1994).  A prime objective is to limit both
statistical and systematic errors in $L_x$ and $T$, in order to obtain
an accurate assessment of the intrinsic dispersion about the mean
$L_X-T$ relation.

\begin{table*}
\begin{minipage}{160mm}
\caption[ ]{Basic data for the 24 clusters in the sample }
\begin{tabular}{lllllllll}
\hline
Name&Redshift&$L_{\rm X}$ {\tiny  [2-10~keV]}& Temperature & Abund. & $L_{\rm
bolom.}$ & $\beta $ & Core radius & Ref.\\
    & & (ergs/s) &(keV) & (rel.  solar) & (ergs/s) & & (kpc)&\\
\hline
Virgo&0.0038&$2.00~10^{43}  $&$2.20 \pm
0.02$&0.44&$6.81~10^{43}$&0.46&12.6&29,29,7,7\\
A262&0.0164&$3.70~10^{43}$&$2.41 \pm 0.05$&0.55& $1.15~10^{44}$&$0.53 \pm 0.03$
&$95 \pm 25$&34,34,12,19\\
A644&0.0704&$1.06~10^{45}$&$6.59 \pm
0.17$&0.33&$2.24~10^{45}$&0.70&204&16,16,20,20\\
A754&0.0534&$1.16~10^{45}$&$7.57 \pm 0.31$&0.23&$2.46~10^{45}$&-&-&1,1\\
A1367&0.0215&$9.09~10^{43}$&$3.50 \pm 0.18$&0.31&$2.26~10^{44}$&$0.50 \pm
0.10$&
$390_{-130}^{+210}$ &16,16,29,20\\
A1656 &0.0232&$1.13~10^{45}$&$8.21 \pm 0.09$&0.22&$2.41~10^{45}$&$0.75 \pm
0.03$&$420 \pm
24$&18,18,8,8\\
A2256&0.0601&$1.11~10^{45}$&$7.51 \pm 0.19$&0.28&$2.35~10^{45}$&$0.81 \pm
0.01$&$542 \pm
12$ &16,16,9,9\\
A2319&0.0564&$2.24~10^{45}$&$9.12 \pm 0.15$&0.24&$4.83~10^{45}$&$0.68 \pm
0.05$&$410 \pm
50$&34,34,19,19\\
A2634&0.0321&$7.09~10^{43}$&$3.07 \pm
0.11$&0.38&$1.89~10^{44}$&0.58&320&1,1,11,11\\
A3526 &0.0109&$7.00~10^{43}$&$3.54 \pm 0.13$&0.54&$1.70~10^{44}$&$0.45 \pm
0.03$&$93 \pm
19$& 32,33,23,23\\
A3921&0.0940&$5.20~10^{44}$&$4.90 \pm 0.55$&0.26&$1.15~10^{45}$&$0.75 \pm
0.04$&$385 \pm
42$&2,2,2,2\\
AWM4&0.0315&$2.53~10^{43}$&$2.43 \pm 0.28$&0.93&$7.52~10^{43}$&0.43
&190&31,31,31,31 \\
AWM7&0.0179&$1.70~10^{44}$&$3.82 \pm 0.15$&0.49&$3.99~10^{44}$
&$0.53 \pm 0.01$&$102 \pm 5$&31,31,28,28\\
Ophui.&0.0280&$1.90~10^{45}$&$9.10 \pm 0.30$&0.24&$4.10~10^{45}$&$0.62
^{+0.01}_{-0.02}$&
$177^{+46}_{-93}$&21,21,24,24\\
&&&&&&&&\\
A665&0.1820&$1.50~10^{45}$&$8.26 \pm 0.90$&0.49&$3.16~10^{45}$&$0.66 \pm 0.05$
&$379 \pm  70$&17,17,5,5\\
A1413&0.1427&$1.82~10^{45}$&$8.85 \pm 0.50$&0.19&$3.92~10^{45}$&0.62
&156&16,16,11,11\\
A2163&0.2010&$6.00~10^{45}$&$14.6 \pm 0.85$&0.40&$1.45~10^{46}$&
$0.62 \pm 0.02$&$305 \pm 19$&14,14,14,14\\
A2218&0.1710&$9.40~10^{44}$&$6.70 \pm 0.50$&0.20&$2.00~10^{45}$&$0.65
^{+0.08}_{-0.05}$&$226^{+80}_{-50}$&25,25,6,6\\
&&&&&&&&\\
A370&0.3700&$1.30~10^{45}$~{\tiny  [1]}&$8.80 \pm
0.80$&0.50&$2.76~10^{45}$&-&-&4,4 \\
A399&0.0715&$7.76~10^{44}$~{\tiny  [3]}&$7.40 \pm
0.50$&0.25&$1.64~10^{45}$&$0.52 \pm 0.05$&$215 \pm
35$&13,15,19,19\\
A401&0.0748&$1.47~10^{45}$~{\tiny  [3]}&$8.40 \pm 0.50$&0.21&$3.15~10^{45}$
&$0.61 \pm 0.01$&$285 \pm 10$&13,15,19,19\\
A1060&0.0114&$2.50~10^{43}$~{\tiny  [2]}&$3.10 \pm 0.20$&0.40&$6.60~10^{43}$
&0.61 &94.&13,30,22,22\\
A3558&0.0478&$4.23~10^{44}$~{\tiny  [2]}&$5.50 \pm 0.25$&-&$9.06~10^{44}$
&0.61 &324&13,27,3,3 \\
Trian.&0.0510&$1.26~10^{45}$~{\tiny  [2]}&$10.3 \pm
0.80$&0.26&$2.77~10^{45}$&$0.63 \pm 0.02$&$286 \pm
16$&13,26,26,26\\
\hline
\end{tabular}
\smallskip

 Notes: column (9) references for the data (respectively X--ray
luminosity, temperature , $\beta$ and core radius):
1.  Arnaud (1994);
2.  Arnaud et al.  (1997);
3.  Bardelli et al.  (1996);
4.  Bautz et al.  (1994); 5.  Birkinshaw, Hughes \& Arnaud (1991);
6. Birkinshaw \& Hughes (1994);
7. B\"ohringer et al. (1994);
8. Briel, Henry \&  B\"ohringer (1992);
9. Briel  \&  Henry (1994);
10. Buote  \& Canizares (1996);
11. Cirimele, Nesci  \& Trevese (1997);
12. David,   Jones \& Forman (1995);
13. Edge et al. (1990);
14. Elbaz, Arnaud \& B\"ohringer (1995);
15. Fujita et al. (1996);
16. Hatsukade (1989);
17. Hughes \& Tanaka (1992);
18. Hughes et al. (1993);
19. Jones \&  Forman (1984);
20. Jones \&  Forman (1990, priv. comm.);
21. Kafuku et al. (1992);
22. Loewenstein  \& Mushotzky (1996);
23. Matilsky, Jones \& Forman (1985);
24. Matsuzawa et al. (1996);
25. McHardy et al. (1990);
26. Markevitch, Sarazin \& Irwin (1996);
27. Markevitch  \& Vikhlinin (1997);
28. Neumann \&  B\"ohringer (1995);
29. Takano (1990) and Koyama (1997, priv. comm.);
30. Tamara et al. (1996);
31. Tsuru (1993);
32. Yamanaka \& Fukazawa (priv. comm. in Fukazawa et al. (1994));
33. Yamashita  1992;
34. Yamashita (1997, priv. comm.)
\end{minipage}
\end{table*}

\subsection{Selection Criteria}

Inclusion of a cluster in the sample is based on three criteria : (i)
small statistical errors in measured temperature $\Delta T/T \!  \lta
\!  10\%$ (at $90\%$ confidence), (ii) weak or
absent cooling flow $\mdot \!  \le \!  100 \msol \yr^{-1}$, and (iii)
$ kT \ge 2 \kev$.  The value of $\mdot$ is set by the relative
importance of core luminosity to the total, and is typically evaluated
using models of multi--phase, steady--state accretion (Thomas, Fabian
\& Nulsen 1987).  Since nearly two--thirds (33/51) of the clusters in
the \xray flux limited sample of Edge \etal (1990) have flows
below this limit (Fabian \etal 1994), our analysis is of a majority
population.

Table~1 presents a listing of the 24 clusters satisfying the above
criteria.  Eighteen have both $T$ and $L_X$ determined by the GINGA
satellite (references are given in the table).  Luminosities for three
nearby clusters (Virgo, in particular) are calculated from scanning
data, which accounts directly for the extended emission outside the
$1.1 \times 2$ sq deg FOV (FWHM) of the satellite instrument.
Temperature measurements for the other six clusters come from ASCA,
with luminosities from {\sl Einstein\/} MPC (3), EXOSAT (2), and ASCA
(1).

We believe this list to be complete with respect to publications
available end of 1996.  Sixteen clusters in our sample are
included in the X-ray flux limited sample of Edge \etal (1990).  Two
more clusters (A2163 and A1413) have fluxes above the Edge et al flux
limit ($1.7 \times 10^{-11} \cgsflux $).  Our sample is $\sim 70\%$
complete as compared to the Edge \etal sample down to a flux of $3.4
\times 10^{-11} \cgsflux$ (considering from now on only weak cooling
flow clusters in the Edge sample for consistency) and highly
incomplete below this flux (3 clusters among 20).  The other clusters
of our sample below the Edge \etal flux limit are either distant
luminous clusters (A3921, A665, A2218, A370) or poor clusters (A2634
and AWM4).  The temperature distribution of our sample shows a clear
deficit of clusters at intermediate temperature ($4-6$ keV) as
compared to the temperature distribution of the Edge \etal sample.

Included in Table~1 are the measured temperature and luminosity in the
2--10 keV energy band (cluster rest frame and corrected for
absorption).  We convert the 2--10 keV luminosity to bolometric
luminosities using an isothermal plasma emission model (Mewe,
Gronenshild \& van den Oord 1985; Mewe, Lemen \& van den Oord 1986)
and the measured temperatures and abundances.  The bolometric
correction factor depends mostly on the temperature; it is higher at
low temperature ($\sim 3.4$ at 2.2 keV), decreases with increasing
temperature up to about 8 keV ($\sim 2.1$) and increases again above
($\sim 2.4$ at 15 keV).  The statistical errors on the luminosities
are always much smaller than the corresponding errors on the
temperature and are thus not taken into account here.

\subsection{Systematic Errors}

To estimate accurately the intrinsic dispersion of the $L_X - T$
relation requires an attempt to understand and quantify sources of
systematic errors.  Such errors can arise from calibration
uncertainties, systematic errors in background subtraction and, for
luminosities deduced from collimated instruments (EXOSAT, GINGA),
improper correction for cluster extent (loss of efficiency at large
radii, emission outside the FOV).  Further uncertainties due to the
uncertainties on the plasma emission model are negligible for clusters
in the considered temperature range (see Arnaud \etal 1991; Arnaud
\etal 1992).

We attempt to guage the magnitude of systematic uncertainty in the
bolometric luminosity by comparing to values derived from the ROSAT
All--Sky Survey (RASS) by Ebeling \etal~(1996) for Abell Clusters.
Eigthteen of the clusters in our sample are in this dataset.  We
convert the RASS band--limited, total unabsorbed flux, derived from a
procedure described in the above reference, to bolometric $L_X$ using
the same plasma emission model as that used on our own data.  In
addition, we include RASS data for Virgo from B\"ohringer \etal~(1994)
and ROSAT imaging data for Triangulum (Markevitch, Sarazin \& Irwin,
1996) and Ophiuchus (Buote \& Tsai 1996), from which we calculate
total luminosities by extrapolating the $\beta$--model fits given in
the references.  Conversion to a bolometric measure for imaging data
is the same as for the RASS data.

We find good agreement between the GINGA and ROSAT bolometric 
luminosities; 16 objects yield a mean ratio of $1.00$ with standard 
deviation of $0.19$.  The agreement is not as good for the five 
non--GINGA members of our sample, but the difference is driven 
entirely by two clusters --- Triangulum and A3558 --- for which the 
ROSAT luminosities are larger than the EXOSAT values we use by factors 
of $1.72$ and $1.56$, respectively.  Both clusters are nearby and have 
extent comparable or larger than the EXOSAT FOV ($45\arcmin \times 
45\arcmin$ FWHM), the virial radius (defined below) of these 2 
clusters are respectively at 49 and 38 arcmin from the center.  Their 
flux can be underestimated due to the neglect of the loss of 
efficiency at large radii.  Furthermore, both observations were 
significantly offset, the pointing offset being specially important 
for A3558 (38 arcmin, Edge 1989).  Although the fluxes published by 
Edge et al.  (1990) include correction for this effect (correction as 
for a point source), this is likely to induce further uncertainties in 
the flux estimate.  For the sake of consistency with the remaining 
dataset, we applied no correction to the published values.

The comparison of the GINGA and ROSAT luminosities, which constitutes
the majority of our sample, indicates that the scatter expected from
systematic uncertainty in $L_X$ is not larger than $20\%$ (rms).
Clusters with large angular extent --- in particular, larger than
GINGA's $1.1\deg \times 2\deg $ field of view --- are likely to incur
larger error, but the number of such objects in our sample is small.
Furthermore, we looked for systematic variation with redshift of the
ratio between the GINGA and ROSAT luminosities.  Such a dependence
with redshift was noticed by Ebeling (1993) for the ratio between the
luminosities (HEAO, MPC and EXOSAT data) given by Henry \& Arnaud
(1991) and the RASS luminosities.  The observed increase of this ratio
with redshift (up to $z \sim 0.07$) was interpreted as being due to an
underestimate of flux for nearby, extended clusters by earlier 
experiments.  This effect is not apparent in the GINGA data, due to 
the larger instrument FOV and the use of scanning mode for the most 
extended clusters.

We now consider systematic errors on temperature.  A detailed estimate
of this error, due to calibration uncertainties and background
subtraction, was done by Hughes \etal (1993) for the GINGA observation
of Coma.  They found a systematic uncertainty of $\pm 0.13$ keV which,
in fact, dominates the especially small statistical uncertainty ($\pm
0.09 \kev$) for this cluster.  Arnaud \etal (1992) also found a
typical systematic uncertainty due to background subtraction of $0.14$
keV for A2163 ($T \se 14.5$ keV).  Although this systematic error is
expected to depend on cluster flux and temperature, these typical
figures show that, for most of the clusters considered here,
statistical errors still dominate the systematic errors.  This is
further confirmed if one compares the temperature estimates from
various instruments.  Generally, good agreement between MPC, EXOSAT
and GINGA estimates is found (David \etal 1993), as well as between
ASCA and other high-energy instruments (Markevitch \& Vikhlinin 1997).
A direct comparison between the temperature estimate for the 7
clusters in our sample with both GINGA and EXOSAT measurements gives a
mean ratio of $0.97$ with standard deviation of $0.06$.

\section{The $L_X - T$ Relation and Implications }

\subsection{The $L_X - T$ Relation}

Figure~1 shows the luminosity--temperature relation for the sample,
with $90\%$ confidence errors plotted for $T$.  Statistical errors on
$L_X$ are typically smaller than the plotted point size and so are not
shown.  The homogeneous GINGA sample data are shown as filled circles,
the remainder as open circles.  The set of clusters define a
remarkably tight relation.  The main outliers are A1060 and
Triangulum, which are anomalously dim (or hot) in comparison to the
others.  The deviation of Triangulum is significantly reduced if the
ROSAT luminosity is employed in place of the EXOSAT value.  As noted
above, this difference is likely due to its large angular extent.  The
peculiarity of A1060 was noted by Loewenstein \& Mushotzky (1996) in a
direct comparison to AWM7, a cluster of similar temperature which is
also in our sample.
 Girardi \etal (1997) also pointed out the departure of this cluster
 from the general $L_X - T$ relation.  They noted that its temperature
 is high compared to its velocity dispersion and suggested some
 anomalies in the dynamical state of the gas.  However they assumed a
 temperature of 3.9 keV from David et al (1993), which is significantly
 higher than the more precise ASCA value adopted here.  From the
 velocity dispersion, $\sigma = 633 $km/s, determined by Girardi \etal
 (1997) and the best fit $\sigma-T$ relation established by Girardi
 \etal (1996), we actually expect a temperature of 2.8 keV, in
 reasonable agreement with the ASCA value of $3.1\pm0.2$ keV.  The
 deviation of A1060 is only sligtly decreased if this value is used
 instead of the ASCA value.  Either both the galaxies and the gas are
 in a very particular dynamical state or the gas structure or content
 in this cluster is abnormal.

\begin{figure}
\vskip -1.2 truecm
\epsfxsize=11.0cm
\epsfysize=11.0cm

\hbox{\hskip -1.0 truecm \epsfbox{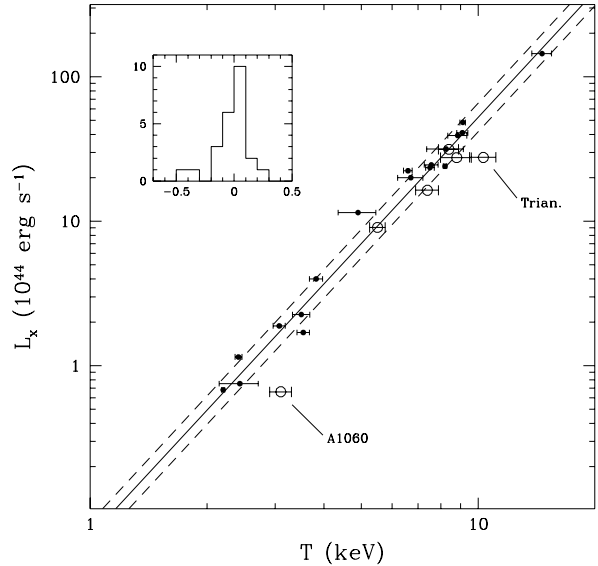} }

\vskip -1.8 truecm
\caption{
The luminosity---temperature relation for the 24 clusters used in
this analysis.  Filled circles are the homogeneous GINGA subset
while open circles denote clusters with ASCA temperatures and
EXOSAT, MPC, or ASCA luminosities.  Error bars plotted on $T$
are $90\%$ confidence limits.   The solid line is the
best--fit relation; the dashed lines are offset
by an amount equal to the {\sl rms\/} deviation in $\logLx$
about the mean relation.
The upper insert shows a histogram of the luminosity residuals
(in dex), which makes apparent the two principal outliers,
A1060 and Triangulum.
}
\label{LxTplot}
\end{figure}

The data are well fit by a power law $\logLx = (2.88 \pm 0.15) \logTd6 
+ (45.06 \pm 0.03)$.  The central values and uncertainties in the fit 
parameters are derived from simple least squares fits (in log space) 
to a large number of bootstrap resamplings employing the estimated $T$ 
measurement errors (quoted errors in slope and intercept are $1 
\sigma$).  Note that since bootstrap procedure is used to estimate 
errors the exact fit method is of secondary importance.  The slope of 
$2.88 \pm 0.15$ is consistent with that found by Ebeling (1993) from 
RASS luminosities ($2.73 \pm 0.39$) but is significantly smaller than 
the value of 3.36 found by David \etal (1993) using mostly MPC and 
EXOSAT data.  This is a direct consequence of the underestimate of 
luminosities for nearby clusters by these last two experiments (see 
above); the observed clusters with low luminosities being on average 
at smaller distance than more luminous clusters which can be observed 
up to larger distances.  Our slope is also consistent at better than 
the $2 \sigma$ level with the recent value of $2.64 \pm 0.16$ 
determined by Markevitch (1998) from analysis of ASCA temperatures and 
ROSAT luminosities of nearby clusters.  Allen \& Fabian (1998) find a 
slope of $2.9 \pm 0.3$ in the non cooling flow portion of their 
combined ASCA/ROSAT sample.

The raw scatter in decimal $\logLx$ about the luminosity--temperature
relation is $0.14$, and some of the variance arises from measurement
errors in temperature.  We estimate this contribution to be $0.06$,
and thus calculate the intrinsic scatter in $\logLx$ at fixed
temperature to be $0.13$, equivalent to a fractional deviation
$\langle (\delta L_X / L_X)^2 \rangle^{1/2} \se 0.30$.  For the
homogeneous GINGA subset of 18 clusters, the fractional deviation
reduces substantially, to $0.17$.

Considering A1060 as an outlier, the fact that there is no significant
trend toward larger scatter in $L_X$ at low temperatures implies that
cooling flows do not contribute substantially to the total luminosity
of these systems.  Our selection criterion of a fixed cooling flow
mass deposition rate is roughly equivalent to an absolute limit in 
core luminosity excess attributed to the cooling flow region.  This 
absolute limit represents a larger fraction of the total luminosity 
from smaller systems, leading to the concern that cooling flows of 
variable strength (but under our $100 \msol \yr^{-1}$ limit) might 
induce larger variance at low temperatures if the cooling region 
dominated the luminosity.  This effect is not apparent in this data 
set of modest number.

\subsection{Limits on Gas Fraction Variations}

The small scatter in Figure~1 supports the notion that at least this
sub--population of weak cooling flow clusters is quite regular in its
structural properties.  In particular, the data appear at odds with
the idea of large gas fraction variations within the virial regions of
clusters, an idea hinted at by the comparison of A1060 to AWM7 by
Loewenstein \& Mushotzky (1996).  We now address this issue using a
spherically symmetric model as a framework for calculation.  The
purpose of the exercise is to provide a context for connecting the
scatter in gas fraction at fixed temperature with the observed
variance in \xray luminosity.  The exercise is non-trivial because
luminosity scatter can be due to changes in gas density structure as
well as overall gas fraction variations.

Consider a cluster with smooth, spherically symmetric distributions of
gas and total mass about its center.  Let $\Mdeltac$ be the total mass
contained in a sphere (of radius $\rdeltac$ about the center) which
encompasses a mean density $\delta_c \rhocrit$, where $\rhocrit \equiv
3\Ho^2/8\pi G$ is the critical density of the universe within which
the cluster is embedded.  The bolometric \xray luminosity of a cluster
is
\begin{equation}
L_X \ = \ \int d^3r \ \rho_{gas}^2(r) \ \Lambda(T(r))
\label{Lxraw}
\end{equation}
 where $\Lambda(T(r))$ is an appropriately normalized emissivity
dependent only on temperature.  In practice, the integral extends not
to infinity, but to a finite radius denoting the edge of the relaxed,
virialized portion cluster.  Spherical collapse models and numerical
simulations suggest that this edge is located at density contrast
$\delta_c \ssimeq 200$, or radius $r_{200}$.

Writing the gas density in terms
of the natural radial variable $y \sequiv r/\rdeltac$, we introduce
the structure function $h(y)$
\begin{equation}
\rho_{gas}(y \rdeltac) \ \equiv \ f_{gas} \ \deltac \, \rhocrit \ h(y) ,
\label{rhogas}
\end{equation}
where the explicit use of the gas fraction within density contrast
$\delta_c$
\begin{equation}
\fgas \ \equiv \ \frac{M_{gas}(<\rdeltac)}{\Mdeltac}
\label{fgasdefn}
\end{equation}
sets the normalization of the structure
function $h(y)$ via the condition
$3 \int_0^1 dy y^2 h(y) \se 1$ .
For an ensemble of clusters, we imagine a set of gas fraction
values $\{ \fgas \}$ and structure functions $ \{ h(y) \}$.

The luminosity in equation (\ref{Lxraw}) with this notation becomes
\begin{equation}
L_X \ \equiv \ \fgas^2 \ Q(\deltac, T)
\label{Lxmod}
\end{equation}
where
\begin{equation}
Q(\deltac,T) \ =  \ \int d^3y \ h^2(y) \ [\Lambda(T(y))
\rhocrit^2  \deltac^2 \rdeltac^3]
\label{Qdefn}
\end{equation}
is largely determined by the structure function $h(y)$, with
weaker dependence on the form of the temperature distribution through
$T(y)$.

The variance in luminosity at fixed temperature has contributions from
gas fraction variations and structural variations
\begin{equation}
\delta_L^2 \ = \ 4 \delta_f^2 + 4 \delta_f \delta_Q + \delta_Q^2 ,
\label{Lxvar}
\end{equation}
where the $\delta$'s are fractional variations at fixed temperature
\begin{equation}
\delta_L \equiv \biggl(\frac {\delta L_X}{L_X}\biggr)_T \ , \
\delta_f \equiv \biggl(\frac {\delta \fgas}{\fgas}\biggr)_T \ , \
\delta_Q \equiv \biggl(\frac {\delta Q}{Q}\biggr)_T  .
\label{deltadefn}
\end{equation}

The magnitude of intracluster gas variations $\delta_f$ can be
determined from the observed scatter in luminosity $\delta_L$
\begin{equation}
\delta_f^2 \ = \ \frac{1}{4} (\delta_L^2 - \delta_Q^2) - \delta_f \delta_Q
\label{delta_f2}
\end{equation}
with an additional assumption about how changes in gas fraction
and cluster structure are correlated.
We assume here that there are no correlations between changes in
gas content and variations in internal structure, meaning
\begin{equation}
\langle \delta_f \delta_Q \rangle \ = \ 0
\label{struct_assump}
\end{equation}
with the angle brackets denoting an ensemble average.

With this assumption, the fact that $\delta_Q^2$ is positive definite leads
to an upper limit on gas fraction variations
from equation~(\ref{Lxvar})
\begin{equation}
\langle \delta_f^2 \rangle^{1/2} \ \le \ \frac{1}{2} \
\langle \delta_L^2 \rangle^{1/2} .
\label{fgaslimit}
\end{equation}
and the resultant values from Figure~1 are
\begin{equation}
\langle \delta_f^2 \rangle^{1/2} \ \le \ \cases{ 0.15 & full sample, \cr
                       0.09 & GINGA subsample. \cr }
\label{fgasvalues}
\end{equation}

Enlarging these limits is possible if an anti--correlation exists
between $\delta_f$ and $\delta_Q$.  Such an anti--correlation would
require gas poor clusters to be more centrally concentrated than
average, and vice--versa.  In this way, the change in luminosity due
to a lower (higher) gas mass would be compensated by an increased
(decreased) central density.  We show below that the observational
data do not support such an anti--correlation.  In addition, one
mechanism capable of generating gas loss from clusters has the
opposite effect.  Heating due to winds from early--type galaxies leads
to a less centrally concentrated ICM structure and slightly depressed
gas fractions within the virial radius (Metzler \& Evrard 1997).  This
implies $\langle \delta_f \delta_Q \rangle > 0$, a result which serves
to strengthen the limit imposed by equation~(\ref{delta_f2}).

We conclude that variations in gas fraction within the virial regions
(overdensities $\deltac \ssim 200$) of weak, cooling flow clusters are
likely to be quite small.  Fractional deviations at fixed temperature
$\delta_f$ may have standard deviation smaller than $10 \%$,
particularly if galactic wind feedback, or some other,
non--gravitational, entropy generating mechanism, has produced
correlated variations in gas fraction and density structure.

The limit on gas fraction variations applies to clusters of similar
temperature; there remains the possibility that the mean cluster gas
fraction varies, perhaps considerably, over the range of temperatures
probed in our sample.

\section{Temperature Dependence of ICM Structure }

In this section, we examine our sample for evidence of variation in
the mean cluster ICM structure and virial gas fraction with
cluster temperature.  We  also discuss the origin of the
$\LxT$ relation slope.

\subsection{The slope of the $\LxT$ relation}

Suppose that cluster density profiles $h_T(y)$ and gas fractions
$\fgas(T)$ are, in the mean, temperature dependent.  If we further
suppose that clusters are internally isothermal, then
equation~(\ref{Lxmod}) can be rewritten in a form which makes clear
the dependence on the shape function $h_T(y)$
\begin{equation}
L_X(T) \ = \ f_{gas}^2(T) \ [\Mdeltac(T) \rhocrit \deltac \Lambda(T)] \
\hat{Q}(T)
\label{Lxmod2}
\end{equation}
with
 $\hat{Q}(T) = (3/4\pi) \int d^3y \, h_T^2(y)$
is the dimensionless emission measure.  Note that $\hat{Q}(T)$ is equal to
$\langle{\rho_{gas}^{2}}\rangle/\langle{\rho_{gas}}\rangle^{2}$, with
the angle brackets denoting the average over the cluster atmosphere.
It is thus a structure factor which depends solely on the gas density
shape and characterises the concentration of the gas distribution.

 The traditional approach is to employ the following set of additional
 assumptions : (i) pure bremsstrahlung emission ($\Lambda(T) \spropto
 T^{1/2}$), (ii) virial equilibrium ($\Mdeltac \spropto T^{3/2}$),
 (iii) structurally identical clusters ($\hat{Q}(T) \se C_1$), and
 (iv) constant gas fraction ($\fgas(T) \se C_2$).  This leads to an
 expected scaling relation between luminosity and temperature of slope
 two
\begin{equation}
L_X(T) \ \propto \ T^2 .
\label{LxTnaive}
\end{equation}

The fact that the slope of the observed relation is significantly
steeper than two implies that one or more of these assumptions does
not hold in the real cluster population.  For example, if we retain
bremsstrahlung emission and the condition of virial equilibrium, then
the observed slope of the \LxT relation leads to a constraint on
structure factor and gas fraction dependence
\begin{equation}
\fgas^2(T) \hat{Q}(T) \ \propto \ T^{0.88 \pm 0.15} .
\label{fgasQcond}
\end{equation}

 The assumption of pure bremsstrahlung is quite accurate above about 2
 keV, but at lower temperatures, line emission becomes significant
 unless the metallicity is very low.  Pure bremsstrahlung emission
 itself does not exactly scales as $T^{1/2}$ due to the additional
 Gaunt factor.  However, this factors varies by less than $4\%$ in the
 1--10 keV temperature range.  On the other hand, the line emission
 reaches about $18\%$ of the total
 emission at 2 keV for a typical abundance of 1/3 the solar value.
 This extra line emission tends to flatten the slope of the $\LxT$
 relation, as low $T$ clusters are boosted in luminosity more than
 their high $T$ counterparts.  This effect is amplified because low
 $T$ clusters tend to have higher abundances.  Given the measured
 abundances we computed $\Lambda(T)$ for our sample (the correct value
 is used in all the following).  We found that the departure from the
 $T^{1/2}$ law is actually small: the best fit slope of the power law
 is 0.36 instead of 0.5.

The assumption of virial equilibrium is supported by cosmological,
gasdynamic simulations (Bryan \& Norman 1998 and references therein).
The relation holds particularly well for $\deltac \simeq
200-2500$, with 200 the fiducial choice of virial radius and 2500
approaching the onset of the cluster core.  We use the
virial theorem as one method for estimating total cluster masses below.

There remains the issue of structural regularity.  Recent numerical
investigation of dark matter halos evolved within hierarchical
clustering cosmogonies indicates that low mass halos are, on average,
more centrally concentrated than high mass halos (Navarro \etal 1997).
Although the magnitude of the change is likely to be small within the
mass range probed by the \xray clusters in our sample, the qualitative
direction is again to flatten the slope of the $\LxT$ relation, if one
assumes that the gas density follows that of the dark matter.

The assumption that the gas traces the dark matter is reasonable if
shock heating through gravitational infall is the only entropy
generating mechanism, but this assumption breaks down if extra energy
sources or sinks are available to the gas.  In particular, if galactic
winds play an important role in heating the ICM, then the extra
entropy dumped into the gas can change its density structure
dramatically (Metzler \& Evrard 1994).  The net result of winds is to
inflate the gas distribution, decreasing the $\hat{Q}$ factor
and causing diminished gas densities and, correspondingly, lowered gas
fractions, at radii interior to the virial radius.  Because
supernova--driven winds impart roughly a fixed specific energy into
their surroundings, their impact is felt more strongly in shallower
potential wells.  This steepens and lowers the intercept of the $\LxT$
relation compared to models without winds.

The magnitude of this effect will depend on details of the assumed
wind model, but plausible models exist.  The numerical simulations of
Metzler (1994) produce a slope of $2.96 \pm 0.05$ in the bolometric
$\LxT$ relation.  Cavaliere, Menci \& Tozzi (1997) create a
semi--analytic model which predicts a progressive steepening of the
$\LxT$ slope toward low temperature clusters, consistent with the data
of Ponman \etal (1996).  Finally, the constant central entropy model
of Evrard \& Henry (1991) predicts a slope of $2.75$ which is
consistent, within the errors, with our observational sample
behaviour.

\subsection{Gas density profiles from \xray imaging}

Azimuthally averaged profiles of cluster \xray surface brightness
images are generally well fit by  the standard, \betamodel form
\begin{equation}
I(r) \ = \ I_0 \, \left[1+(r/r_c)^2\right]^{-3\beta+1/2}
\label{betaeq}
\end{equation}
derived by assuming an isothermal gas with density distribution
$\rho_{gas}(r) = \rho_0 (1+(r/r_c)^2)^{-3\beta/2}$.
Fits to this model exist for all but two (A754 and A370) of the members
of our sample.  Table~1 lists values from the literature
of the best--fit core radius $r_c$ and outer slope parameter
$\beta$.  Errors on these parameters are also given in the table when
available.

\begin{figure}

\vskip -0.5 truecm
\epsfxsize=11.0cm
\hbox{\hskip -1.0 truecm
\epsfbox{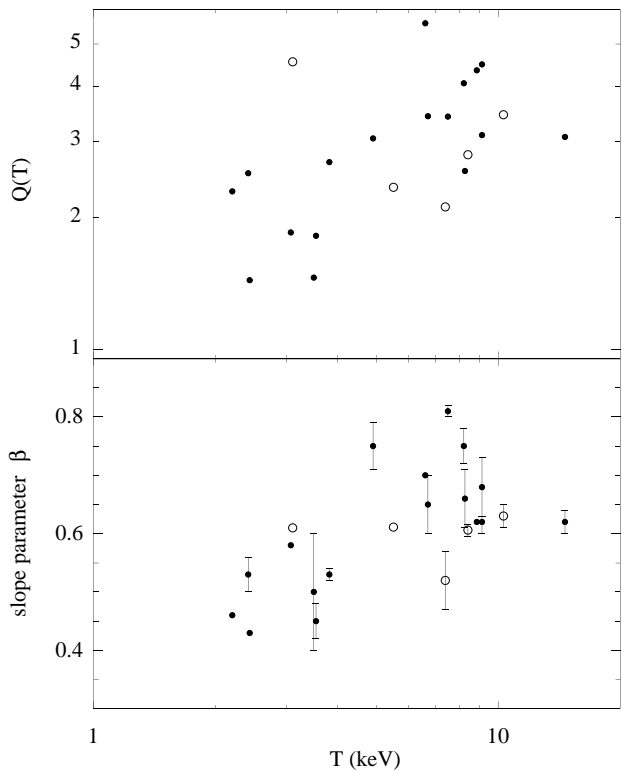} } \vskip -4.8 truecm

\caption{ Values of the gas outer slope parameter $\beta$ (bottom)
and structure factor $\hat{Q}$ (top), derived from \betamodel
fits to \xray images plotted against temperature for 22 of the 24
members of our sample.  Point styles are the same as the previous
figures.  Error bars available in the literature are shown.  }
\label{betaplot}
\end{figure}

Figure~\ref{betaplot} displays the best fit slope parameters against
temperature.  As noted previously (David et al.  1990; White
1991; Mohr \& Evrard 1997), there is a clear trend toward decreasing
$\beta$ toward lower $T$.  This trend is not likely to be an artifact
of the fitting procedure, as Mohr \& Evrard (1997) see a similar trend
with $\beta$ defined in a non--parametric, non--azimuthally averaged
fashion.

The heterogeneous nature and lack of error estimates for nearly a
third of these data make it difficult to determine the formal
statistical significance of this effect.  As a reasonable estimate, we
calculate the bi--weight modes (Beers, Flynn \& Gebhardt 1990) for the
subsamples of 8 clusters with $T<4 \kev$ and 14 clusters with $T>4
\kev$.  We applied a $10\%$ fractional error to each data point ---
slightly larger than the \rms fractional error of $7.8 \%$ for the 15
quoted error values --- and determined the expected subsample means
and standard error from bootstrap analysis.  The values of
$\langle\beta\rangle_{lo} \se 0.51 \pm 0.03$ and
$\langle\beta\rangle_{hi} \se 0.66 \pm 0.03$ for the low-- and
high--$T$ subsamples differ at the $3.3\sigma$ level.  We suspect that
future analysis with a more regular dataset will confirm the
statistical significance of this trend.

The implication is that the ICM density structure of clusters is
temperature dependent, and the sense of the effect is such that
$\hat{Q}(T)$ is an increasing function of temperature (see
Figure~\ref{betaplot}).  Models of cluster formation incorporating
galactic winds exhibit a trend of $\beta$ versus $T$ similar to that
seen in the observational data (Metzler \& Evrard 1997).  The trend of
$\hat{Q}(T)$ with temperature is in the right direction to steepen the
\LxT relation, but whether it is entirely responsible for this effect
requires investigation of the trend of mean gas fraction with
temperature.

\subsection{Intracluster Gas Fraction Values}

The determination of absolute values of the gas fraction requires
a method to estimate the total cluster mass.
We show here that two different methods for total mass estimation
lead to different conclusions about the behavior of the gas fraction
in clusters.

\subsubsection{Mass determination methods}

Estimating cluster total masses has traditionally been done with
the ``$\beta$--model'' (BM) approach (Cavaliere \& Fusco-Femiano 1976),
for which the total mass is
\begin{equation}
M_{\beta}(r) \ = \ 1.13 \times 10^{15} \beta \ \frac{T}{10 \kev}
\ \frac{r}{\rm Mpc} \ \frac {(r/r_c)^2} {1+(r/r_c)^2} \msol
\label{betamass}
\end{equation}
This model assumes a non-rotating, isothermal gas distribution
in hydrostatic equilibrium, with gas density profile of the form assumed
to derive the surface brightness image, equation~(\ref{betaeq}).
This leads to a binding mass which increases linearly with radius
outside the core.  Estimating the mass at some density threshold
$\deltac$ is readily done by calculating the mean interior density
as a function of radius.  Outside the core region, this radius scales as
\begin{equation}
r_{\delta_c,\beta}  = \ 2.78 \ \beta^{1/2}
\biggl( \frac {\deltac} {500} \biggr)^{-1/2}
\biggl( \frac {T} {10 \kev} \biggr)^{1/2} h_{50}^{-1} \mpc
\label{rbeta}
\end{equation}

An alternative method is to employ the virial theorem (VT) at
fixed density contrast, which leads to
\begin{equation}
M_{VT}(\rdeltac) \ = \ M_{10}(\delta_c) \
\biggl( \frac{T}{10 \kev} \biggl)^{3/2}.
\label{VTmass}
\end{equation}
Calibration of this relation with numerical experiments by Evrard,
Metzler \& Navarro~(1996, hereafter EMN) shows that the intercept
$M_{10}(\deltac)$ is remarkably insensitive to assumptions regarding
the background cosmology and to the effects of galactic winds.  The
experiments show that the VT method has intrinsically smaller variance
compared to the BM approach, because the imaging information
introduces an additional source of noise.  A recent intercomparison of
12 cosmological gas dynamic techniques lends credence to these
calibrations.  Frenk \etal (1998) find agreement within $4\%$ for the
total cluster mass and $6\%$ for the mass
weighted cluster temperature among codes varying substantially in
algorithmic detail, resolving power and numerical parameterization.

At density contrast $\deltac \se 500$,  EMN find
$M_{10}(500) \se 2.2 \times 10^{15} \hfiftyinv \msol$, with a
corresponding physical size scaling as
$\rfiveh(T) \se 2.5 (T/10 \kev)^{1/2} \hfiftyinv \mpc$.
At $\deltac \se 200$, the mass intercept is
$M_{10}(200) \se 2.9 \times 10^{15} \hfiftyinv \msol$ and
the physical size is $r_{200}(T) \se 3.7 \hfiftyinv \mpc$, roughly $50\%$
larger than $\rfiveh$.  All values are quoted for the current epoch,
and both scale with redshift as $(1+z)^{-3/2}$.  Though typically quite
small, this redshift scaling is taken into account in our analysis.
We examine gas fractions at these two
characteristic density contrasts below.  For the majority of imaging data,
$\rfiveh$ is comparable to or smaller than the radius used to
fit the gas model parameters while values at $r_{200}$ require a
modest extrapolation outside the \xray imaged region in nearly all
cases.

To determine the gas mass, we employ values of
$\beta$ and $r_c$ from Table~1, and determine the central
gas density by normalizing to the bolometric luminosity, assuming the
latter is due to emission within $r_{200}$.
Though not exact, the typical error introduced by the
use of $r_{200}$ as a cutoff is small, since for most clusters the
contribution to the luminosity outside this radius is negligible.  As a 
check, we normalized the gas mass using $\rfiveh$ --- an extreme 
choice since many cluster images extend beyond this radius and since 
$\rfiveh$ is nearly a factor 2 smaller than $r_{200}$ --- and the 
fractional increase in gas density was typically $\sims 5\%$,
with a maximum fractional increase of $24\%$ (absolute increase
$\sims 3\%$) for low $\beta$ clusters.  As detailed below, our
procedure produces a population mean gas fraction at $\rfiveh$
for hot clusters consistent with the determination of Evrard (1997)
from the observational samples of White \& Fabian (1995) and
David \etal (1995).

\subsubsection{Gas fraction estimates}

\begin{figure}
\vskip -0.6 truecm
\epsfxsize=10.0cm
\epsfysize=10.0cm

\hbox{\hskip -0.5 truecm \epsfbox{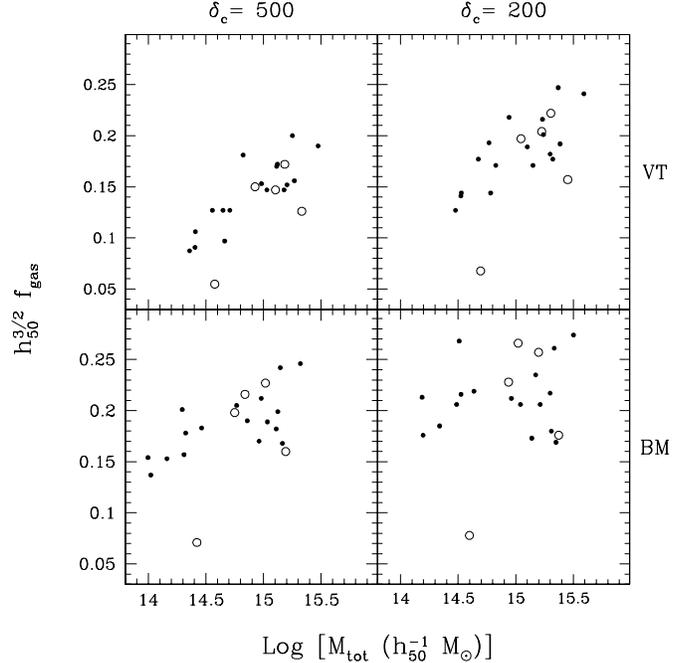} }

\vskip -0.1 truecm
\caption{
Gas fraction versus total mass at inferred density contrasts
$\deltac \se 500$ and $200$ derived using the VT (upper) and BM
(lower) approaches discussed in the text.  Symbol types are
identical to those in Figure~1.
}
\label{Massplot}
\end{figure}

Gas fractions and total masses derived from the VT and BM methods
at $\deltac \se 500$ and 200 are shown in Figure~\ref{Massplot}.
There are trends apparent which reflect the different
behavior of the two mass estimate methods to the input data,
particularly the sensitivity to the slope parameter $\beta$.
Comparing equations~(\ref{betamass}) and
(\ref{VTmass}) at $\rfiveh$, and ignoring $(r_c/\rfiveh)^2$, one finds
\begin{equation}
\frac {M_{\beta}(\rfiveh)} {M_{VT}(\rfiveh)}
\ = \ \biggl( \frac {\beta} {0.79} \biggr)^{3/2} .
\label{massrat}
\end{equation}
From Figure~\ref{betaplot}, all but one of the values of $\beta$ in
our sample lie below $0.79$.  The \betamodel estimates are
therefore consistently lower than those of the virial theorem, and
this is clearly reflected in the overall horizontal shift of the
data points between the upper and lower rows.  Furthermore,
the systematic trend of $\beta$ with temperature means the discrepancy
is larger at smaller temperatures/masses.
This effect also causes the gas fractions at constant $\deltac$ in the
BM to be consistently higher than those of the VT,
 although the difference is small at $\delta=200$.

In some of the panels, a trend of mean gas fraction with cluster mass
or temperature is evident to the eye.  The lack of rigorous,
homogeneous, statistical error estimates for the $r_c$ and $\beta$
values in our sample makes it difficult to make precise statements
regarding the significance of these trends.  We proceed with an
approximate treatment in which we assign a $1\sigma$ fractional
uncertainty of $30\%$ to each of the gas fraction estimates.  This
approach allows us to explore trends in the data and provides an
estimate of their statistical significance under an assumed error
budget.  The $30\%$
error may be generous given that $T$, $r_c$ and $\beta$ are typically
determined to $\ssim 5\%$ accuracy, and the fact that gas mass errors
are smaller than errors in $\beta$ and $r_c$ alone, because these
parameters are tightly correlated.  However, the ``cosmic variance''
in the binding mass estimates is believed to be $\sim 15-30\%$ at
these density contrasts (EMN), and this represents the limiting 
accuracy of any dataset.

\begin{figure}
\vskip -4.1 truecm
\epsfxsize=14.0cm
\epsfysize=14.0cm

\hbox{\hskip -1.4 truecm \epsfbox{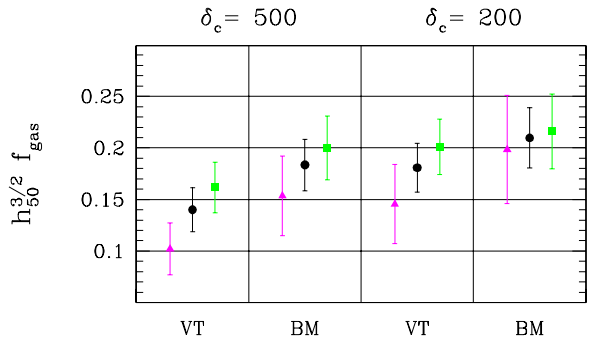} }

\vskip -5.4 truecm
\caption{
Bi--weight mean gas fractions at density contrasts of 500 and 200
for the VT and BM methods.  Data for the full sample of 22 clusters
are shown as filled circles, the low $T$ subsample of 8 clusters as
triangles, and the high $T$ subsample of 14 clusters as filled
squares.  Error bars are $90\%$ confidence limits
($1.65 \sigma$) determined from bootstrap resampling and
assuming a fixed fractional error of $30\%$ on individual cluster
gas fraction estimates.
}
\label{fbarplot}
\end{figure}

Bi--weight mean gas fractions determined from a bootstrap analysis
assuming fixed $30\%$ fractional errors in each measurement are
listed in Table~2 and summarized in Figure~\ref{fbarplot}.
Data for the whole sample as well as cool and hot subsamples are
provided.  Uncertainties quoted in the Table are standard errors
of the mean, while $90\%$ confidence errors (assuming Gaussian
statistics) are plotted in the Figure.

\begin{table}
\caption{Bi--weight Mean Gas Fractions (\%, $h_{50}^{-3/2}$) \label{fgastable}}
\begin{flushleft}
\begin{tabular}{lcc}
 & \hskip 2.5truecm $\deltac \se 500$ & \hskip 2.0truecm $\deltac \se 200$ \\
\end{tabular}
\begin{tabular}{lcccc}
Sample  & VT & BM & VT & BM \\
\hline
All & $14.0 \pm 1.3$ & $18.3 \pm 1.5 $ & $18.1 \pm 1.4$  & $21.0 \pm 1.8$ \\
$T<4 \kev$ & $10.2 \pm 1.5$ & $15.3 \pm 2.4 $ & $14.5 \pm 2.1$ & $19.9 \pm
3.2$ \\
$T>4 \kev$ & $16.2 \pm 1.5$ & $20.0 \pm 1.9 $ & $20.1 \pm 1.6$ & $21.6 \pm
2.2 $ \\
\end{tabular}
\end{flushleft}
\end{table}

The trend of rising gas fraction with increasing radius (lower
$\deltac$) is apparent with both mass estimation methods, and reflects
the more extended nature of the gas density compared to the assumed or
inferred dark matter profiles.  Gas fraction values are consistently
higher in the BM approach, due to the smaller, inferred total masses
noted above.  As a result, significantly different sample means can be
quoted depending on the mass method and extent of radial coverage of
the dataset.  Compare, for example, $\fbargas \se 0.140 \pm 0.013$ for
the VT at $\deltac \se 500$ with $\fbargas \se 0.210 \pm 0.018 $
for the BM at $\deltac \se 200$.

The range becomes wider when one considers subsets of the data defined
by temperature.  All the panels in Figure~(\ref{fbarplot}) display a
trend of decreasing gas fraction at lower $T$, but only the VT method
at $\deltac \se 500$ is deemed statistically significantly ($2.8
\sigma$) under our $30\%$ error hypothesis.  Due to the
robustness of the bi--weight mean estimator, this conclusion is valid
with or without the inclusion of the principal outlier A1060.  It is
perhaps also worth noting that the discrepancy between A1060 and the
mean of the cool subsample is smaller under the VT approach.

The value of $16.2 \pm 1.5\%$ for the hot subsample in the VT case
is consistent with the $95\%$ confidence region
$\bar{f}_{gas}(\rfiveh) \se 17.0 \pm 1.7\%$ ($\hfifty \se 1$)
determined by Evrard (1997) from the combined samples of White \&
Fabian (1995) and David, Jones \& Forman (1995).  Only 3 of the 26
clusters in the combined study have $T < 4 \kev$ (and these have
fairly uncertain gas fractions which make negligible contribution to
the sample mean), so comparison with our high--$T$ subsample is
reasonable.  The agreement between these two studies is not entirely
trivial.  Although there is nearly $50\%$ overlap with our subsample
--- 6 of our 14 are in common with White \& Fabian --- we employ here
different temperature estimates in many cases (though all are
consistent within the quoted errors) and we normalize the gas mass
differently.

A less obvious trend in the data is the larger variance in gas
fractions in the BM case compared to the VT. This is more apparent in
the $T$--dependent subsamples, since the trend of mean gas fraction
with temperature mimics variance in the full VT sample.  The lower
scatter in the VT method confirms theoretical predictions that the
$\beta$ parameter acts as an extra source of noise in the BM mass
estimates (EMN).  An increase in scatter is also seen at larger
radii/lower densities, consistent with the numerical model predictions
and interpreted simply as reflecting the longer relaxation timescales
in the outer regions of clusters.

\subsection{Implications}

There is strong evidence from the slope of the $\LxT$ relation and
from the behavior of the image fall-off parameter $\beta$ that cluster
structure varies systematically with temperature.  Whether the virial
gas fraction varies with cluster temperature is a question whose
answer is both model and scale dependent.  The model dependency we
highlight is the choice of total mass estimation method, while the
chosen density contrast sets the scale.  For example, the \betamodel
applied at $\delta_c \se 200$ produces similar mean values in high and
low temperature subsamples --- $0.216 \pm 0.022$ and $0.199 \pm
0.032$, respectively --- whereas the virial theorem at $\deltac \se
500$ yields significantly different values in these subsamples ---
$0.162 \pm 0.015$ against $0.102 \pm 0.015$, respectively (see Table~2
and Figure~\ref{fbarplot}, all values scale as $\hfifty^{-3/2}$).

That the implications of the observed luminosity--temperature relation 
are model dependent can be seen by considering the behavior of terms with 
temperature dependence contained in the defining relation, 
equation~(\ref{Lxmod2}).  The structure factor $\hat{Q}(T)$ appears to 
increase with temperature, but this effect is not solely responsible 
for the steepening of the $\LxT$ relation.  The dependence with $T$ of 
 the other factors depends on the total mass estimation method.  In 
the VT approach, while $\Mdeltac(T)$ scales classically as $T^{3/2}$, 
$f_{gas}(T)$ increases with $T$.  In the BM case, while $f_{gas}$ is 
found to be nearly constant, the slope of the $\Mdeltac(T)-kT$ 
relation is steeper than $3/2$, due to the extra $\beta$ factor and 
its increase with $T$ (equation~(\ref{massrat})).  Both cases result 
in a steepening of the $f_{gas}^2(T) \ \Mdeltac(T)$ term to a slope 
greater than the canonical $3/2$.  The temperature dependence of this 
term contributes about equally with $\hat{Q}(T)$ to the steepening of 
the $\LxT$ relation.

Allan \& Fabian (1998) take a different approach in analysis of
a cluster dataset combining ASCA spectroscopy with ROSAT imaging.
They fit ASCA data to a multi-component spectral model which includes
emission from a central cooling flow (Johnstone \etal 1992), absorption from 
galactic hydrogen, intrinsic hydrogen absorption, and a superimposed 
background plasma at temperature $T$.  Compared to simpler fits
without cooling flow and intrinsic absorption, the best fit
temperatures in their complex fits are higher, and the best fit 
slope of the $\LxT$ relation is shallower, $2.33 \pm 0.43$.  Allen \& 
Fabian emphasize that this slope is consistent with the
straightforward theoretical expectation of $2$;  however, note that the
range of slopes allowed in our fit does overlap the error range
quoted above (derived from a change $\delta \chi^2 = 2.71$ in their
$\chi^2$ analysis).  

The approximate nature of our error analysis limits the conclusions
which can be drawn regarding gas fraction variations.  A definitive
analysis awaits uniform treatment of a large, homogeneous dataset.
There is some supporting evidence for deficient gas fractions in poor
groups (Hickson, 1997 and references therein),
 but there also appears to be larger scatter at low temperatures (\eg,
 Ponman \etal 1996) and observational uncertainties on the ICM mass
 remain large in these systems.  In particular the separation of the
 ICM emission from the X--ray emission linked to individual galaxies
 is ambiguous and difficult with present instruments (Mahdavi \etal
 1997, Mulchey and Zabludoff 1998).

\section{Concluding Discussion}

The narrow scatter in the $\LxT$ relation of weak cooling flow
clusters ($\dot{M} \le 100 \msol \yr^{-1}$) indicates that the
structure of the intracluster medium in this majority sub--class is
very regular.  Under a conservative assumption of no correlation
between virial gas fraction and internal, structural variations
($\langle \delta_Q \delta_f \rangle \se 0$), we place upper limits on
the {\sl rms\/} percentage variation in virial gas fraction of $15\%$
for the entire sample of 24 clusters, and $9\%$ for the
homogeneous GINGA subset of 18 clusters.

These limits are conservative because the \xray imaging data supports
the point of view that gas in low $T$ clusters is more extended than
that of high $T$ clusters (Figure~\ref{betaplot}).  Mohr \& Evrard
(1997) find the same result with a non--parametric analysis of \xray
images.  They also derive limits on gas fraction variations similar to
those found here based on the tightness of the \xray size--temperature
(ST) relation.  The agreement is non--trivial because the ST analysis
is essentially {\sl local}, being based on an differential isophotal
area, while the $\LxT$ relation analysis is {\sl global}, based an the
integrated emission of the entire cluster atmosphere.  Both analyses
are consistent with idea that energy input from galactic winds has led
to extended gas atmospheres in low $T$ clusters (David, Forman \&
Jones 1990,1991; White 1991; Metzler \& Evrard 1997; Cavaliere \etal
1997; Ponman \etal 1996).

In wind models, gas loss is correlated with structural extension of
the gas, leading to $\langle \delta_Q \delta_f \rangle > 0$.  The
inferred variance in gas fraction derived from
equation~(\ref{delta_f2}) could be substantially smaller than the
limits phrased above if winds have played an important evolutionary
role.  Numerical simulations of clusters used by Mohr \& Evrard (1997) exhibit
less than 5\% fractional variation in their virial gas fractions, and
a similar value is seen in models with galactic winds (Metzler \&
Evrard 1997).  The expected ``cosmic variance'' in virial gas
fractions is thus quite small.

Comparison of a number of different simulation techniques paints a
similar picture.  Frenk \etal (1998) find agreement at the 5\% level
($1 \sigma$ deviation) for the virial gas fraction of a single cluster
determined by twelve independent gas dynamic methods.  This indicates
that systematic uncertainties in the tested numerical simulation
techniques are small for the simplest case of a single phase ICM.  
However, it is important to note that the quoted simulations ignore 
galaxy formation and its attendant multi--phase, gas structure.  
Introducing galaxy formation via cooling and star formation is likely 
to increase the scatter in ICM gas fraction values, and it will be 
interesting to see whether viable models can do so without violating 
the observational limits presented here.

Evidence for trends in $\fbargas$ with cluster temperature is
ambiguous.  A key uncertainty lies in estimation of the total, binding
mass.  The \betamodel approach produces masses which are
characteristically smaller than the virial theorem method as currently
calibrated by numerical simulations.  For a cluster with $\beta \se
0.6$, the masses inferred at $\delta_c \se 500$ differ substantially,
$M_\beta \se 0.66 M_{VT}$.  This different behavior, coupled with the
temperature dependence of $\beta$, leads to differing conclusions on
whether the mean gas fraction varies with cluster temperature.

For the same reason there is some ambiguity in the origin of the
steepening of the $\LxT$ relation.  The fact that the gas is more
concentrated in high-kT cluster explains at least half of the
effect, directly through the structure factor in the X--ray luminosity.
In the \betamodel approach, the temperature dependence of the ICM
structure can fully account for the $\LxT$ relation slope, via the
additional dependence of the total mass with $\beta$.  On the other 
hand, in the VT approach, the gas mass fraction increases with $T$ and 
this effect contributes about equally to the $\LxT$ relation slope 
increase.  Both effects, lower gas fraction and inflated gas 
distribution in low T systems, are expected in models incorporating 
the effects of galactic winds.

 Independent methods for estimating virial masses are crucial, and
weak gravitational lensing provides an important approach.  Allen
(1997) provides evidence that weak lensing and \betamodel mass
estimates are consistent, but the present errors at radii between
$r_{500}$ or $r_{200}$ are too large to discriminate between the BM
and VT approaches.  We suggest that weak lensing analysis of a small
sample of moderate redshift clusters selected to have $\beta \lta 0.6$
from \xray imaging could settle the score, if individual measurement
uncertainties could be kept below a few tens of percent at $\delta_c
\se 500$.

Future \xray missions which provide simultaneous imaging and
spectroscopic capability (AXAF, XMM, Astro-E) will provide
improvements over the current dataset.  The ideal analysis would be
one of a flux limited sample with central, cooling flow regions
excised (Markevitch 1998), using a nearly homogeneous dataset of
temperatures, luminosities and surface brightness maps derived from
the same telescope.  Such data would minimize variance caused by
instrumental uncertainties and open the door to very high precision
measurements of intracluster gas fractions.

Finally, small scatter in virial gas fractions lends support to
arguments against an Einstein--deSitter universe based on cluster
baryon fractions and primordial nucleosynthesis (White \etal 1993;
Evrard 1997).  The standard counterargument to this idea has been that
the physics of the ICM is more complicated than is assumed in most
analytic or numerical models.  However, if additional physics beyond
gravity, shock heating and galactic wind input is at work in cluster
ICM atmospheres, then the extra processes involved must conspire to
keep the scatter in the observed $\LxT$ and ST relations small.  Since
$L_X$ and $R_I$ (the ``size'' in the ST relation) are manifestly
different measures (global {\sl vs.\/} local), it is not obvious that
interesting physical mechanisms, such as a strongly multi--phase ICM
(Gunn \& Thomas 1996), can be added in such a way as to preserve the
current agreement between observations and current models
incorporating the effects of galactic winds on the ICM.

\section*{Acknowledgments}
We would like to thank Dr Koyama for useful correspondence on the
GINGA observations of Virgo and Dr Yamashita for providing us with the
GINGA results on A262 and A2319.  We thank Alain Blanchard for a 
careful reading of the manuscript.  This work was supported by NASA
through Grant NAG5-2790 and by the CIES and CNRS at the Institut
d'Astrophysique in Paris.  AEE is grateful to the members and staff of
the IAP for the kind hospitality extended during a sabbatical visit.


\begin{thebibliography}{}
\bibitem[ ] { } Allen S.W., 1997, \mnras, in press, astro-ph/9710217

\bibitem[ ] { } Allen S.W. \& Fabian, A.C. 1998, \mnras, 297, L57. 

\bibitem[ ] { } Arnaud M., Lachieze-Rey M., Rothenflug R., Yamashita
K., Hastukade I, 1991, \aa, 243, 56

\bibitem[ ] { } Arnaud M., Hughes J.P., Forman W., Jones C.,
Lachieze Rey M., Yamashita K., Hastukade I, 1992, \apj , 390, 345

\bibitem[ ] { } Arnaud, M., 1994, in Cosmological Aspects of X-ray
Clusters of Galaxies, W.C. Seitter ed., NATO ASI Series, vol 441 p.197

\bibitem[ ] { } Arnaud  M., Rothenflug  R., B\"ohringer H., Neumann
D., Yamashita, K., 1997, \aa, submitted

\bibitem[ ] { } Bardelli S., Zucca E., Malizia A., Zamorani  G.,
Scaramella R., Vettolani  G.,  1996, \aa , 305, 435

\bibitem[ ] { } Bautz M.W., Mushotzky  R., Fabian  A.C., Yamashita  K.,
Gendreau K.C., Arnaud K.A., Crew G.B., Tawara Y.,  1994, \pasj , 46,
L131

\bibitem[ ] { } Beers T.C., Flynn K., Gebhardt K., 1990,
\aj, 100, 32

\bibitem[ ] { } Birkinshaw M., Hughes J.P., Arnaud K.A.,  1991,
\apj , 379, 466

\bibitem[ ] { } Birkinshaw M., Hughes J.P.,  1994, \apj , 420, 33

\bibitem[ ] { } B\"ohringer H., Briel U.G., Schwarz R.A., Voges W.,
Hartner G., Tr\"umper J., 1994, Nat, 368, 828

\bibitem[ ] { } Briel U.G., Henry J.P., B\"ohringer H.,  1992, \aa 259,
L31

\bibitem[ ] { } Briel U.G., Henry J.P.,  1994, \nature , 372, 439

\bibitem[ ] { } Bryan, G.L. \& Norman, M.L. 1998, \apj, 495, 80.

\bibitem[ ] { } Buote D.A., Canizares C.R.,  1996, \apj , 457, 565

\bibitem[ ] { } Buote D.A., Tsai J.C.,  1996, \apj , 458, 27

\bibitem[ ] { } Cavaliere A., Fusco-Femiano R., 1976, \aa, 49, 137

\bibitem[ ] { } Cavaliere A.,  Menci N., Tozzi P., 1997,
\apj, 484, L21

\bibitem[ ] { } Cirimele G., Nesci R., Trevese D., 1997 ,
\apj , 475, 11

\bibitem[ ] { } David L.P., Arnaud K.A., Forman W., Jones C., 1990,
\apj, 356, 32

\bibitem[ ] { } David L.P., Forman W., Jones C., 1990, \apj, 359, 29

\bibitem[ ] { } David L.P.,  Forman W., Jones C., 1991,
\apj, 380, 39

\bibitem[ ] { }David L.P., Slyz A., Jones C., Forman W., Vrtilek
S.D., Arnaud K.A., 1993, \apj, 412, 479.

\bibitem[ ] { }David L.P., Jones C., Forman W., 1995, \apj, 445, 578.

\bibitem[ ] { } Ebeling H., 1993, phD thesis, MPE

\bibitem[ ] { } Ebeling H., Voges W., B\"ohringer H., Edge A.C.,
Huchra J.P., Briel J.G., 1996, \mnras, 281, 799

\bibitem[ ] { } Edge A.C., 1989, phD Thesis, Leicester Univ.

\bibitem[ ] { } Edge A.C., Stewart G.C., Fabian A.C.J,  Arnaud K.A.,
1990, \mnras, 245, 559

\bibitem[ ] { } Edge A.C., Stewart G.C., 1991, \mnras, 252, 414

\bibitem[ ] { } Elbaz D., Arnaud M., B\"oringer H., 1995, \aa , 293,
337

\bibitem[ ] { } Evrard A.E., Henry J.P., 1991, \apj, 383, 95

\bibitem[ ] { } Evrard A.E., Metzler C.A., Navarro J.F., 1996,
\apj, 469, 494.

\bibitem[ ] { } Evrard, A.E., 1997, \mnras, 292, 289.

\bibitem[ ] { } Fabian A.C., Crawford C.S., Edge A.C., Mushotzky R.F.,
1994, \mnras, 267, 779

\bibitem[ ] { } Frenk C.S., White S.D.M. \etal 1998, \apj, submitted.

\bibitem[ ] { } Fujita Y., Koyama K., Tsuru T., Matsumoto H.,  1996,
\pasj , 48, 191

\bibitem[ ] { } Garasi C., Burns J.O. \& Loken C. 1997, BAAS, 191, 5308.

\bibitem[ ] { } Girardi M., Fadda D., Giuricin G., Mardirossian F.,
Mezzeti M., Biviano A., 1996, \apj, 457, 61

\bibitem[ ] { } Girardi M., Escalera E., Fadda D., Giuricin G.,
Mardirossian F.,
Mezzeti M.,  1997, \apj, 482, 41

\bibitem[ ] { } Gunn K.F., Thomas P.A., 1996, \mnras, 281, 1133

\bibitem[ ] { } Hatsukade I.,  1989, phD Thesis, ISAS RN 435

\bibitem[ ] { } Henry J.P., Arnaud K.A., 1991, \apj , 372, 410

\bibitem[ ] { } Hickson, P., 1997, \araa , 35, 357

\bibitem[ ] { } Hughes J.P., Tanaka Y., 1992, \apj, 398, 62

\bibitem[ ] { } Hughes J.P., Butchler J.A., Stewart G.C., Tanaka Y.,
1993, \apj, 404, 611

\bibitem[ ] { }Johnstone, R.M., Fabian, A.C., Edge, A.C. \& Thomas,
P.A. 1992, \mnras, 267, 779.

\bibitem[ ] { }Jones C., Forman, W.,  1984,  \apj , 276, 38

\bibitem[ ] { } Jones C., Forman, W.,  1990, private communication

\bibitem[ ] { } Kafuku S., Yamauchi M., Hattori H., Kawai N.,
Matsuoka M.,  1992, in Frontiers of X-ray astronomy, ed.  Y.  Tanaka,
K.  Koyama (Universal Academay Press, Tokyo), P.483

\bibitem[ ] { } Loewenstein M., Mushotzky R.F.,  1996, \apj , 471, L83

\bibitem[ ] { } Mahdavi A., B\"ohringer H., Geller M., Ramella M. , 1997, \apj
, 483, 68

\bibitem[ ] { } Matilsky T., Jone, C., Forman W.,  1985, \apj , 291,
621

\bibitem[ ] { } Matsuzawa H., Matsuoka M., Ikebe Y., Mihara T.,
Yamashita K.,  1996, \pasj , 48, 565

\bibitem[ ] { } McHardy I.M., Stewart G.C., Edge A.C., Cooke B.,
Yamashita K., Hatsukade I., 1990, \mnras, 242, 215

\bibitem[ ] { } Markevitch M.L., Sarazin C.L., Irwin J.A.,  1996, \apj
, 472, L17

\bibitem[ ] { } Markevitch M., Vikhlinin A., 1997, \apj 474, 84

\bibitem[ ] { } Markevitch M., 1998, \apj submitted, astro-ph/9802059

\bibitem[ ] { } Metzler C.A., 1994, in Clusters of Galaxies, eds. F.
Durret, A. Mazure \& J. Tr\^an Thanh V\^an (Editions Frontieres,
Gif--sur--Yvette), p. 251.

\bibitem[ ] { } Metzler C.A., Evrard A.E., 1994, \apj, 437, 564

\bibitem[ ] { } Metzler C.A., Evrard A.E., 1997, \apj, submitted,
astro-ph/9710324

\bibitem [ ] { } Mewe  R., Gronenshild  E.  H.  B.  M., van den Oord  H.
J., 1985,
\aa,  62, 197

\bibitem[ ] { } Mewe  R., Lemen  J.R., van den Oord  H.  J., 1986, \aa, 65,
511

\bibitem[ ] { }Mitchell R.J., Ives J.C., Culhane J.L., 1977, \mnras, 181, 25p

\bibitem[ ] { }Mitchell R.J., Dickens R.J., Bell Burnell S.J., Culhane
J.L., 1979,
\mnras, 189, 329


\bibitem[ ] { }Mohr J.J., Evrard A.E.,  1997, \apj, 1997, 491, 38

\bibitem[ ] { }Mushotzky R.F.,Serlemitsos P.J., Smith B.W., Boldt E.A.,
Holt S.S., \apj, 225, 21

\bibitem[ ] { } Mulchey J.S., Zabludoff A.I., 1998, \apj, 496, 73

\bibitem[ ] { }Navarro J.F., Frenk C.S., White S.D.M., 1997, \apj, 490, 493

\bibitem[ ] { } Neumann, D.M, B\"ohringer H.,  1995, \aa , 301, 865

\bibitem[ ] { }Ponman T.J., Bourner P.D.J., Ebeling H., B\"ohringer H.A.
1996, \mnras, 283, 690.

\bibitem[ ] { } Takano S., 1990, PhD Thesis, Tokyo Univ.  and Koyama,
K., 1997, private communication

\bibitem[ ] { }Tamara T., Day C.S., Fukazawa Y., Hatsukade I., Ikebe
Y., Makishima K., Mushotzky R.F., Ohashi T., Takenaka K., Yamashita
K., 1996, \pasj , 48, 671

\bibitem[ ] { } Tsuru,T., 1993, phD Thesis, ISAS RN 528

\bibitem[ ] { } Thomas P.A., Fabian A.C., Nulsen P.E.J., 1987, \mnras,
228, 973.

\bibitem[ ] { } White R.E., 1991, \apj, 367, 69

\bibitem{} White S.D.M., Navarro J.F., Evrard A.E., Frenk  C.S., 1993,
Nature, 366, 429



\bibitem[ ] { }White D.A., Fabian A.C., 1995, \mnras, 273, 72.

\bibitem[ ] { } Yamanaka M., Fukazawa Y.,  private communication in
Fukazawa et al., 1994, \pasj , 46, L55

\bibitem[ ] { } Yamashita K., 1992, in Frontiers of X-ray astronomy, ed. Y.
Tanaka, K. Koyama (Universal Academay Press, Tokyo), P.475

\bibitem[ ] { } Yamashita K., 1997, private communication


\end{thebibliography}
\end{document}